\pdfoutput=1

\documentclass[onecolumn,10pt,conference]{IEEEtran}

\usepackage{graphicx}  
\newif\ifpdf
\ifx\pdfoutput\undefined
\pdffalse 
\DeclareGraphicsExtensions{.eps} \else
\pdfoutput=1 
\pdftrue \DeclareGraphicsExtensions{.pdf,.jpg,.png} \fi

\begin{document}

\title{A Physical Model of Wireless Network}  

\author{\authorblockN{Jean-Marc Kelif}
\authorblockA{Orange Labs\\
Issy-Les-Moulineaux, France\\
jeanmarc.kelif@orange-ftgroup.com}
}

%


\maketitle

\begin{abstract}
Using an approach developed in physics, we propose a new framework for the study of cellular networks. The key idea of the physical network model we propose is to replace the discrete base stations (BS) entities by a continuum of transmitters which are spatially distributed in the network. This allows us to establish a closed form formula of the other-cell downlink interference factor $f$, as a function of the location of the mobile. We define here $f$ as the ratio of outer cell received power (i.e. the power received from other cells) to the inner cell received power. 
This physical model allows calculating 
the influence of interference on any mobile in a cell, whatever its position. 
Results obtained with that closed-form formula are close to the ones obtained by simulations using a traditional hexagonal network model. Since the physical model allows to establish a closed form formula of the interference factor, it allows to do analytical studies of wireless networks such as outage probability, quality of service, capacity.
\end{abstract}

\IEEEpeerreviewmaketitle

\section{Introduction}

The estimation of cellular networks capacity is one of the key points before deployment and mainly depends on the characterization of interference. As downlink is often the limited link w.r.t. capacity, we focus on this direction throughout this paper, although the proposed framework can easily be extended to the uplink. In CDMA and OFDMA systems, an important parameter for this characterization is the other-cell interference factor $f$ (OCIF). The precise knowledge of the OCIF allows the derivation of outage probabilities, capacity evaluation and then, the definition of Call Admission Control mechanisms. 

On the downlink, \cite{Vit94} \cite{Vit95} aimed at computing an average OCIF over the cell by numerical integration in hexagonal networks. In \cite{Gil91}, Gilhousen et al. provide Monte Carlo simulations and obtain an histogram of $f$.  In \cite{Ela05}, other-cell interference is given as a function of the distance to the BS thanks to Monte-Carlo simulations. Chan and Hanly \cite{Chan01} precisely approximate the distribution of the other-cell interference. They however provide formulas that are difficult to handle in practice. 

In contrast to previous works in the field, the modelling key of our approach is to consider the discrete BS entities of a cellular network as a continuum. Recently, the authors of \cite{ToTa01} described a network in terms of macroscopic quantities such as the node density. The same idea is used in \cite{Jac04} for ad hoc networks. They however assumed a very high density of nodes in both papers. 

This paper extends the framework proposed in \cite{Kel01}  and \cite{KeCoGo01} that provides a simple closed-form formula for $f$ on the downlink as a function of the distance to the BS, the path-loss exponent, the distance between two BS, and the network size. We show that a network can be considered as a continuum of BS, by using an analogue approach as in physics.


The rest of the paper is organized as follows. In Section \ref{secCellular} we first introduce the interference factor $f$. It characterizes the relative influence of the interference generated by transmissions originating from the other cells of the network.  
In Section \ref{PhysicalBS},
following an analogue approach with the gravitation's case, we show that a transmitting BS can be replaced by a density of BS distributed all over the cell.
In Section \ref{PhysicalNetwork}, we generalize that approach to a network.
Considering the base stations of a network as a continuum, we establish a closed form formula of the interference factor $f$. .
By comparison with simulations, we establish that the physical model and the traditional hexagonal model are two simplifications of the reality. None is a priori better than the other but the latter is widely used, especially for dimensioning purposes. That is the reason why a comparison is useful. We show that the physical model is accurate even when the density of BS is very low (see section \ref{valid})
In Section V we conclude.

\section{Interference model and notations}\label{secCellular}
We consider a cellular radio system and we focus on the downlink. BS have omni-directional antennas, so that a BS covers a single cell. If a mobile $u$ is attached to a station $b$ (or serving BS). The following power quantities are considered:

\begin{itemize}
\item 
$P_{b,u}$ is the transmitted power from station $b$ towards mobile $u$ (for user's traffic);
\item  
$P_{b}= P_{cch}+ \Sigma_{u}P_{b,u}$ is the total power transmitted by station $b$. 
In CDMA systems,  
$P_{cch}$ represents the amount of power used to support broadcast and common control channels.
\item
$p_{b,u}$ is the power received at mobile $u$ from station $b$;
we can write \(p_{b,u} = P_{b}\; g_{b,u}\); $g_{b,u}$ designates the pathloss between station $b$ and mobile $u$
\item
$S_{b,u}=P_{b,u} \: g_{b,u}$ is the useful power received at mobile $u$ 
from station $b$ (for traffic data); 
\end{itemize}

The total amount of power experienced by a mobile station $u$ in a cellular system can be split up into several terms: useful signal ($S_u$), interference and noise ($Noise$). It is common to split the system power into two terms: $I_u=I_{int,u}+I_{ext,u}$, where $I_{int,u}$ is the {\it internal} (or own-cell) received power and $I_{ext,u}$ is the {\it external} (or other-cell) interference. Notice that we made the choice of including the useful signal $S_u$ in $I_{int,u}$, and, as a consequence, it has to be distinguished from the commonly considered own-cell interference.

With the above notations, we define the interference factor in $u$, as the ratio of total power received from other BS to the total power received from the serving BS: $f_u=I_{ext,u}/I_{int,u}$. The quantities $f_u$, $I_{ext,u}$, and $I_{int,u}$ are location dependent and can thus be defined in any location $x$ as long as the serving BS is known. In downlink, a coefficient $\alpha$,  may be introduced to account for the lack of perfect orthogonality in the own cell. 

In this paper, we will use the signal to interference plus noise ratio (SINR) as the criteria of radio quality:$\gamma_u$ is the SINR evaluated at MS $u$ and
$\gamma^*_u$ is the SINR target for the service requested by MS $u$. This figure is a priori different from the SINR evaluated at mobile station $u$. However, we assume perfect power control, so $SINR=\gamma^*_u$ for all users. 

\subsection{CDMA system}
With the introduced notations, the SINR experimented by $u$ can thus be derivated (see e.g. \cite{Lagr05}):
\begin{equation} \label{SIR}
\gamma^*_u = \frac{S_{u}}{\alpha(I_{int,u}-S_{u}) + I_{ext,u} + Noise} 
\end{equation}

From this relation, we can express $S_u$ as:
\begin{equation}
S_u = {\frac{\gamma^*_u }{1+\alpha \gamma^*_u}}
            \;I_{int,u} \;
            (\alpha + I_{ext,u}/I_{int,u} + Noise /I_{int,u} )
\end{equation}
and the transmitted power for MS $u$, $P_{b,u} = S_{u} / g_{b,u}$, using relations  $I_{int,u}=P_{b}g_{b,u}$ and $f_u=I_{ext,u}/I_{int,u} $ as:
\begin{equation} \label{Pbu}
P_{b,u} = {\frac{\gamma^*_u }{1+\alpha \gamma^*_u}}  (\alpha P_{b} + f_uP_{b} + Noise /g_{b,u}).
\end{equation}
From this relation, the output power of BS $b$ can be computed as follows:
\begin{equation}
P_b=P_{cch}+\sum_u P_{b,u}, \nonumber 
\end{equation}
and so, according to Eq.\ref{Pbu},
\begin{equation} \label{Pb}
P_b=\frac{P_{cch}+\sum_u \frac{\gamma^*_u }{1+\alpha \gamma^*_u} \frac{Noise}{g_{b,u}}}{1-\sum_u \frac{\gamma^*_u }{1+\alpha \gamma^*_u} (\alpha + f_u)}. 
\end{equation}

\subsection{OFDMA system}
In OFDMA, the data are multiplexed over a great number of subcarriers. There is no internal interference, so we can consider that $\alpha(I_{int,u}-S_{u})=0 $.
Since $I_{ext,u}=\sum_{j\neq b}P_j g_{j,u}$, the expression (\ref{SIR}) can be written, 
\begin{equation}
\gamma_u = \frac{P_{b,u}g_{b,u}}{\sum_{j\neq b}P_j g_{j,u}  + Noise} 
\end{equation}

so we have
\begin{equation}
\gamma_u = \frac{1}{f_u  + \frac{Noise}{P_{b,u}g_{b,u}}} 
\end{equation}

Since $ \frac{Noise}{P_{b,u}g_{b,u}} << f_u$, typically for cell radii less than about 1 km, we can negligt this term and write

\begin{equation}
\gamma_u = \frac{1}{f_u} 
\end{equation}

Remark

We showed that the interference factor is a key parameter to analyze wireless networks. In the aim to propose a closed form formula of $f$, we first develop a physical model of the network.


\section{Physical Analysis }\label{PhysicalBS}

The physical network model considers that a discrete set of BS of a cellular network can be replaced by a continuum of BS. This approach can be compared to other approaches in physics, where discrete entities are considered as a continuum. For example, to analyze certain types of problems concerning the gravitation, a discrete mass $M$ can be considered as a continuum. And in electrostatics, some discrete charges can be considered as a continuum, too.  Hereafter, we remind these approaches based on the Gauss theorem, and we propose an analogy with a wireless network. We first remind the Gauss theorem, applied to mechanics and electrostatics, which allows to replace a discrete mass (or charge) by a continuum of mass (or charge). Afterwards, writing the local expression of the Gauss theorem, we remind its equivalence with the fact that the gravitational (electrostatic) field can be interpreted as the \textit{response} of the space to a mass (charge) solicitation. We go on, observing that it is also the case for a BS. More precisely the power received at a given point of a wireless network can be interpreted as the response of this point to a BS transmitting power solicitation. And we show that a BS can be replaced by a density of BS, when the power received has the following form  $K r^{- \eta}$    with $\eta =2$. Considering a multidimensional analysis of wireless networks, we show that this results can be generalized for $\eta > 2$. Finally we interpret that last analysis, in the case of a realistic network. 

\subsection{Gauss theorem: global expression}

In a mechanical system, let's consider a mass $m$ inducing a gravitational force at a given point of the space, located at the distance $r$ from the mass. The gravitation is characterized by a function form $\frac{1}{r^2}$. The Gauss theorem shows that the gravitational effect of a non punctual object with a mass density $\rho_m$, a volume $V$ and a mass $m$ can be replaced by an equivalent one with the same mass which is located at its gravity mass centre. For example the Earth can be replaced by a point in its centre where all the Earth mass would be concentrated. The gravitational effects of these two approaches are equivalent. 
In fact the Gauss theorem expresses that a field created by a \textit{non ponctual} object of mass $m$(gravitational case) or charge $q$(electrostatic case) at a given point of a system can be calculated by considering only the distance to the mass centre (or the charge centre), and $not$ the distances to each elementary mass (or charge).
The Gauss theorem allows\textit{ replacing} a mass distribution density (or charge distribution density) by one unique mass (or charge).

\vspace{0.2cm}
\subsubsection{Electrostatic case}

The electric field $\vec{E}$ flow through a closed surface $S$ is proportional to the sum of the charges which are inside this surface.  

\begin{equation} \label{GaussElec}
\int \int_S \vec{E}.d\vec{S}  = \frac{Q}{\epsilon_0} 
\end{equation}
where $Q= \int \int \int_V \rho_c dV $  and $\rho_c $  is the density of charges in the volume V.

\vspace{0.2cm}
\subsubsection{Gravitational case}
The gravitation force $\vec{F}$ flow through a closed surface is equal to the sum of the interior masses $m_i$ multiplied by  $- 4 \pi G$.

\begin{equation} \label{GaussGrav}
\int \int_S \vec{F} . d \vec{S} = -4 \pi G \sum m_i 
\end{equation}
where $\sum m_i = \int \int \int_V \rho_m dV $  and $\rho_m $  is the density of mass in the volume V.
 
\vspace{0.2cm}
\subsubsection{Application to a Wireless Network}
The question is: can we replace a set of discrete transmitting sources (BS) by a continuum of sources ($\rho_{BS}$)? We notice that a source BS is characterized by its transmitting power $P_t$. In other terms, considering the magnetic field $\vec{B}$ and the Poynting vector $\vec{P} = \vec{E} \wedge \vec{B}$ characterizing the transmitting electromagnetic power, is it possible to write such a relation:
 
\begin{equation} \label{GaussElec2}
\int \int_S \vec{P_r} . d \vec{S} =  \sum P_t 
\end{equation}
where $\sum P_t = \int \int \int_V \rho_p dV $  and $\rho_p $  is the density of power in the volume V,  
$P_r$ is the module of the Poynting vector received at a given point, and $P_t$  the power transmitted by a BS?

\noindent
Since a BS is characterized by its transmitting power $P_t$, that question means: is it possible to replace a punctual BS by a BS density all over a cell? 
We can express differently that question as follows: Is the effect of a punctual BS at a given point of a network the same as the effect of a BS density all over a cell (analogy with mechanics or electrostatic)? 
To answer that last question, we will express the Gauss theorem differently. Precedently, we expressed the global (or integral) expression of the Gauss theorem. Hereafter we analyze its local expression. 

\subsection{Gauss theorem: local expression} 
Hereafter, we show that, from the local expression of the Gauss theorem written as the \textit{Poisson} equation, we can interpret the gravitation (and the electric) field as the \textit{Response of the Isotropic Space} to a mass (charge) density solicitation \cite{JPe06}.


The force $\vec{F(r,t)}$ created at each point of a system where there is a density of mass is obtained as the sum of all the infinitesimal contributions  $\rho(r',t)$ 

\begin{equation} \label{GaussGravDelta}
\delta F(\vec{r},t)= G \rho(r',t) \delta^3 r' \frac{\vec{r}-\vec{r'}}{\|r-r'\|^3} 
\end{equation} 
of a continuum distributed matter:
\begin{equation} \label{GaussGravDelta2}
F(\vec{r},t)= \int G \rho(r',t) \delta^3 r' \frac{\vec{r}-\vec{r'}}{\|\vec{r'}-\vec{r'}\|^3} d^3r'.
\end{equation}

Introducing a gravitational potential $\psi(r,t)$: 

\begin{equation} 
\psi(r,t)= -\int G \rho(r',t) \delta^3 r' \frac{1}{\|\vec{r}-\vec{r'}\|} d^3 r' 
\end{equation} 
and when we notice that 

\begin{equation} 
grad(\frac{1}{\|\vec{r}-\vec{r'}\|}) = \frac{\vec{r}-\vec{r'}}{\|\vec{r}-\vec{r'}\|^3}, 
\end{equation} 
we can write  $ F(r,t)= - grad (\psi(r,t))$ , and finally:   
We can write the potential as a convolution: 

\begin{equation} 
\psi(r,t)= - G \rho(r,t) * \frac{1}{\|r\|} . 
\end{equation} 

And so $ grad (\psi(r,t)) = - G grad(\rho(r,t) * \frac{1}{\|r\|})$ . 

The function $\frac{1}{\|r\|})$ is the Green function of the Laplacien $\Delta $, to a constant. We have $\Delta = -4 \pi \delta(r)$ where $\delta(r)$ is the Dirac distribution. So we have  

 \begin{equation} \label{GaussGravLoc}
\Delta (\psi(r,t))= 4 \pi G \rho(r,t)*\delta(r)
\end{equation}
Since the Dirac is the neutral element of the convolution algebra, we finally obtain the Poisson equation:

 \begin{equation} \label{GaussGravLoc2}
\Delta (\psi(r,t))= 4 \pi G \rho(r,t)
\end{equation}

These equations mean that Gravitation can be interpreted as the \textbf{Response of the isotropic Space to a density mass solicitation}.

\subsection*{Concluding remark}
It is equivalent to say that the gravitation is the response of the isotropic space density mass solicitation and to write the equation \ref{GaussGravLoc} which expresses that it exists a gravitational potential $\psi$  whose Laplacien is proportional to a mass density .

\subsection{Equivalent of Poisson Equation for a Base Station }

We consider a BS, and the zone covered by it denoted a cell. The BS transmits the power $P_t$. The power received at a given point of the network can be expressed as proportional to $\frac{P_t}{r^{\eta}}$. We can also say that each base station has an impact to a point of the network given by that expression. 
This effect can be interpreted as the \textbf{Response of the environment to the BS power transmission solicitation}. 

We thus can introduce a \textit{power potential} $\psi_p$, and a density of BS $\rho_{BS}$, and write this response with an equivalent of the relation \ref{GaussGravLoc}  for the case of a BS, as follows:

\begin{equation} \label{GaussPowerLoc}
\Delta (\psi_P(r))= K \rho_{BS}*\delta(r)
\end{equation}

\vspace{0.5cm}

\textbf{Remark}: due to the property of the  Dirac function, a localized BS transmitting power can be expressed as a localized density of BS transmitting power over the transmission surface.

\vspace{0.5cm}
In the case of $\eta=2$, we express the analogy with the gravitation as follows. The power $P_r$ received  at each point of a system is due to a density of transmitting BS, and is obtained as the sum of all the infinitesimal contributions (analogy with (\ref{GaussGravDelta})) of a continuous base station distribution.

\begin{equation} \label{GaussPowerLoc2}
\delta P(r,t)= K \rho_{BS}(r',t) \delta^3 r' \frac{r-r'}{\|r-r'\|^3}    
\end{equation}

As a consequence, for $\eta=2$, we can derive a "Base station's Poisson equation" for the power received at a given point of a network.  In our analysis, a BS can be defined by its transmitting power $P_t$ localized at the center of the cell. It can be written $P_t \delta_{BS}(r)$. Using the Gauss theorem, we can replace a BS by a density of BS  all over the cell area covered by the BS. 
So the impact of a localized BS on the network is the same as the one due to a density of BS distributed all over the cell area. As a consequence the localized BS can be replaced by a BS density all over the cell. 

\subsection{Some concluding remarks}
We recalled that the gravitational (or electrostatic field) can be interpreted as the Response of the isotropic Space to a density mass solicitation (Poisson). Consequently the gravitational effects of a non ponctual object of mass $m$ and density of mass $\rho_m$ are the same as the ones induced by an object with the same mass $m$ concentrated at its mass centre (Gauss). And finally a mass $m$ can be replaced by a volume with a density of mass   (Gauss). 
Following an analogue approach with the gravitation's case, we showed that the power received at a point of a system coming from a BS can be interpreted as the Response of the isotropic Space to a BS solicitation. So we can write a "Base station's Poisson equation". Consequently the impact of the transmit power $P_t$ of a BS (localised at the centre of a cell) is the same as the one induced by a density of BS all over the cell with the same total transmitting power. And finally a transmitting BS can be replaced by a density of BS distributed all over the cell.

We notice that these developments are available due to the fact that we consider a free and isotropic space ($\eta=2$).

\subsection{Wireless network analysis}
For $\eta=2$, we showed that a BS can be replaced by a density of BS. However the parameter $\eta$ is not equal to the value 2 in a real environment. It is generally comprised between 3 and 4. This is due to the reflections, refractions of radio propagation ... all phenomena which transform an isotropic and free environment to a non one. 

\vspace{0.5cm}
\textbf{Remark:}
\noindent
The  equation  \ref{GaussPowerLoc} can be generalized to obtain the
classical definition of the gravitational potential in a D-dimensionnal space \cite{JPe06}. So $\psi$ can be written as proportional to  $ \frac{1}{r^{D-2}}$ in a multidimensional space where $D$ represents the dimension of the space.

\vspace{0.5cm}
\noindent
That property can be interpreted as follows: the response of the system to a solicitation depends on the dimension of the space considered for the system. For example the gravitational force could be written as proportional to $\frac{1}{r^{D-1}}$. Generally, $D$=3 for the "normal" space.
In wireless network, the values of $\eta$ should be  equal to 2 in a free space. The differences observed for the values of $\eta$ can be interpreted as follows. 
When the space is not free, the parameter $\eta$  characterizes a pathloss parameter, for a multidimensional system, with $D>3$, in which the space would be free and isotropic. In this case, the response of the space to the power solicitation of the BS takes into account the environment, as if the set of elements constituting  the environment (trees, cars, building…) would characterize other dimensions effects. 
In other words, since $\eta$ = 3, the 3 dimensional-space can be expressed as a "four" dimensional one. This four dimensional space (D=4) is free and isotropic (in a point of view of power reception), and the response to the power solicitation can be written as proportional to $   \frac{1}{r^{D-1}} = \frac{1}{r^{3}}  $ . 

Another way to express this approach consists in saying that a non-free environment in dimension 3 with  $\eta$= 3 can become free in dimension 4: the environment increases the space dimensions. 
We can apply the Gauss theorem in this 4-dimensional space and replace a  BS (i.e. a localized power transmission) by a continuum of BS all over the surface cell.

We notice that considering a unique value of  $\eta$ means that the environment is isotropic


 \section{A Physical Network Model}\label{PhysicalNetwork}
In this section, we first present the model, derive the closed-form formula for $f$, and compare it through Monte-Carlo simulations in a hexagonal network.
\subsection{OCIF formula} \label{OCIFformula}
We showed that it is possible to replace a ponctual BS (i.e. transmitting power) by a density of BS distributed all over the cell. We generalize that approach to a network. The key modelling step of the model we propose consists in replacing a given fixed finite number of BS by an equivalent density of transmitters which are spatially distributed in the network.
This means that the transmitting power is now considered as a continuum field all over the network. 
In this context, the network is characterised by a MS density $ \rho _{MS}$ and a base station density $ \rho _{BS}$ \cite{Kel01}. We assume that MS and BS are uniformly distributed in the network, so that 
$\rho _{MS}$ and $\rho _{BS}$ are constant. As the network is homogeneous, all base stations have the same output power $P_b$.

\begin{figure}[htbp]
\centering
\includegraphics[scale=0.5]{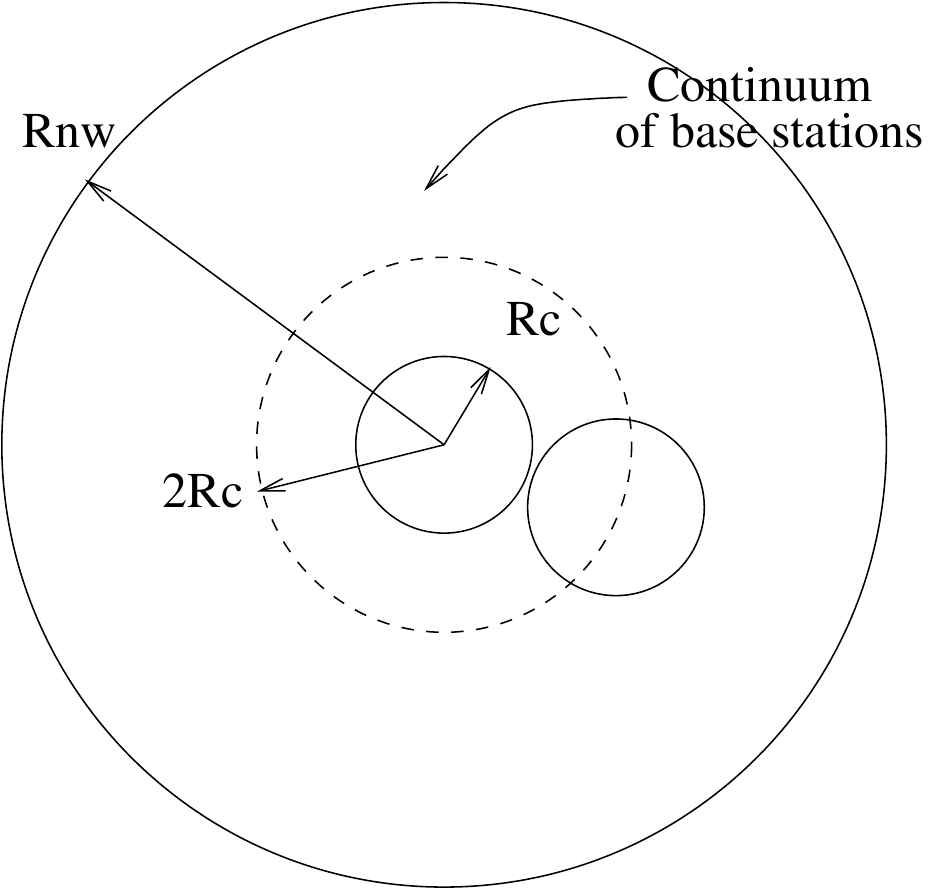}
\caption{Network and cell of interest in the physical model; the distance between two BS is $2R_c$ and the network is made of a continuum of base stations.}
\label{fluidmodel}
\end{figure}

We focus on a given cell and consider a round shaped network around this centre cell with radius $R_{nw}$. The half distance between two base stations is $R_c$ (see Figure \ref{fluidmodel}).

For the assumed omni-directional BS network, we use a propagation model, where the path gain, $g_{b,u}$, only depends on the distance $r$ between the BS $b$ and the MS $u$. The power, $p_{b,u}$, received by a mobile at distance $r_u$ can thus be written $p_{b,u}= P_{b}Kr_u^{-\eta}$, where $K$ is a constant and $\eta>2$ is the path-loss exponent. 

Let's consider a mobile $u$ at a distance $r_u$ from its serving BS. Each elementary surface $zdzd\theta$ at a distance $z$ from $u$ contains $\rho _{BS}zdzd\theta $ base stations which contribute to $I_{ext,u}$. Their contribution to the external interference is $\rho _{BS}zdzd\theta P_bKz^{-\eta}$. We approximate the integration surface by a ring with centre $u$, inner radius $2R_c-r_u$, and outer radius $R_{nw}-r_u$.

\begin{eqnarray}
I_{ext,u}&=&\int_0^{2\pi}\int_{2R_c-r_u}^{R_{nw}-r_u}\rho _{BS}P_bKz^{-\eta}zdzd\theta \nonumber \\
&& \nonumber
\end{eqnarray}
\begin{equation}
=\frac{2\pi \rho _{BS}P_bK}{\eta - 2} \left [ (2R_c-r_u)^{2-\eta}-(R_{nw}-r_u)^{2-\eta} \right ].  
\end{equation}

Moreover, MS $u$ receives internal power from $b$, which is at distance $r_u$: $I_{int,u}=P_bKr_u^{-\eta}$. So, the interference factor $f_u=I_{ext,u}/I_{int,u}$ can be expressed by:

\begin{equation} \label{fnw}
f_u=\frac{2\pi \rho _{BS}r_u^{\eta}}{\eta - 2} \left [ (2R_c-r_u)^{2-\eta}-(R_{nw}-r_u)^{2-\eta} \right ]. 
\end{equation}

Note that $f_u$ does not depend on the BS output power. This is due to the fact that we assumed an homogeneous network and so all base stations emit the same power. In our model, $f$ only depends on the distance $r$ to the BS and can be defined in each location, so that we can write $f$ as a function of $r$, $f(r)$. Thus, if the network is large, i.e. $R_{nw}$ is big in front of $R_c$, $f_u$ can be further approximated by:

\begin{equation} \label{fnwinf}
f(r)= \frac{2\pi \rho _{BS}r^{\eta}}{\eta - 2} (2R_c-r)^{2-\eta}. 
\end{equation}

This closed-form formula will allow us to fastly compute performance parameters of a cellular radio network. We notice that formula can be applied to {\it any wireless system} (TDMA, ad-hoc...) as long as $ \rho _{BS}$ represents the density of {\it interfering transmitting nodes}.

\subsection{Comparison with hexagonal model network } \label{valid}

\begin{figure}[htbp]
\centering
\includegraphics[scale=0.5]{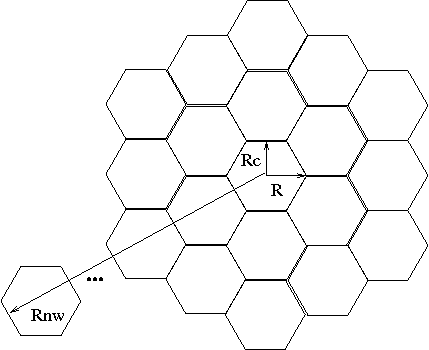}
\caption{Hexagonal network and main parameters of the study.}
\label{hexanw}
\end{figure}

In this section, we aim at comparing the physical approach presented above to an hexagonal classical one. In this perspective, we will compare the figures obtained with Eq.\ref{fnw} with those obtained numerically by simulations. The simulator assumes an homogeneous hexagonal network made of several rings around a centre cell. Figure \ref{hexanw} shows an example of such a network with the main parameters involved in the study.

\begin{figure}
\centering
\includegraphics[scale=0.5]{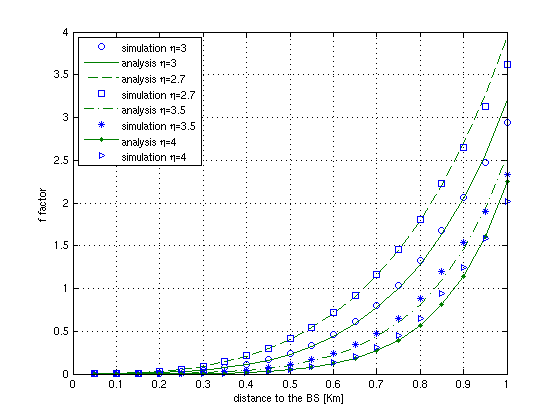}
\caption{Interference factor vs. distance to the BS; comparison of the physical model with simulations on an hexagonal network with $\eta=2.7$, $3$, $3.5$, $4$.}
\label{fcomparison6}
\end{figure}

The comparison is done numerically by computing $f$ in each point of the cell and averaging the values at a given distance from the BS. Figure \ref{fcomparison6} shows the simulated interference factor as a function of the distance to the base station. Simulation parameters are the following: $R=1$~Km, $\alpha=0.7$, $\eta$ between $2.7$ and $4$, $\rho_{BS}=(3\sqrt{3}R^2/2)^{-1}$, the number of rings is 15, and the number of snapshots is 1000. Eq.\ref{fnw} is also plotted for comparison. In all cases, the physical model matches very well the simulations on an hexagonal network for various figures of the path-loss exponent. It allows calculating the influence of a mobile, whatever its position in a cell. 

Note that the considered network size can be finite and chosen to characterize each specific local network's environment (urban or country, macro or micro cells…). 
Since the physical model allows to establish a closed form formula of the interference factor, it allows to do analytical studies of wireless networks such as outage probability, quality of service, capacity \cite{KeCoGo01}.

\section{Conclusion}
We proposed a new framework for the study of cellular networks called the physical network model. 
The key idea of the physical model is to replace the discrete base stations entities by a continuum of transmitters which are spatially distributed in the network. This allows us to obtain simple analytical expressions of the main characteristics of the network, the interference factor $f$. 
We focused on CDMA and OFDMA networks.
Following an analogue approach with the gravitation's case, we showed that the power received at a point of a system coming from a BS can be interpreted as the response of the isotropic space to a BS solicitation. So we can write a "Base station's Poisson equation". Consequently the impact of the transmit power $P_t$ of a BS (localised at the centre of a cell) is the same as the one induced by a density of BS all over the cell with the same total transmitting power. Therefore a transmitting BS can be replaced by a density of BS all over the cell. Generalizing that approach to a whole network, we established a closed form formula of the interference factor. Comparing this network approach to a classical hexagonal one, we showed the physical model and the traditional hexagonal model are two simplifications of the reality. None is a priori better than the other but the latter is widely used, especially for dimensioning purposes. That is the reason why a comparison is useful.
However, since the physical model allows to establish a closed form formula of the interference factor, it allows to do analytical studies of wireless networks such as outage probability, quality of service, capacity.


\begin{thebibliography}{1}

\bibitem{Vit94}
A.~J.~Viterbi, A.~M.~Viterbi, and E.~Zehavi,
Other-Cell Interference in Cellular Power-Controlled CDMA, IEEE Trans. on Communications, Vol. 42, No. 2/3/4, Freb/Mar/Apr. 1994.
\bibitem{Vit95}
A.~J.~Viterbi,
CDMA - Principles of Spread Spectrum Communication, Addison-Wesley, 1995.
\bibitem{Lagr05}
X.~Lagrange,
Principes et \'evolutions de l'UMTS, Hermes, 2005.
\bibitem{Ela05}
S.~E.~Elayoubi and T.~Chahed,
Admission Control in the Downlink of WCDMA/UMTS, Lect. notes comput. sci., Springer, 2005. 
\bibitem{Gil91}
K.~S.~Gilhousen, I.~M.~Jacobs, R.~Padovani, A.~J.~Viterbi, L.~A.~Weaver, and C.~E.~Wheatley,
On the Capacity of Cellular CDMA System, IEEE Trans. on Vehicular Technology, Vol. 40, No. 2, May 1991.
\bibitem{Chan01}
C.~C.~Chan and Hanly,
Calculating the Outage Probabability in CDMA Network with Spatial Poisson Traffic, IEEE Trans. on Vehicular Technology, Vol. 50, No. 1, Jan. 2001.
\bibitem{Kel01}
J.-M.~Kelif, 
Admission Control on Fluid CDMA Networks, Proc. of WiOpt, Apr. 2006.
\bibitem{KeA01}
J-M. Kelif and E. Altman,
Downlink Fluid Model of CDMA Networks, Proc. of IEEE VTC Spring, May 2005.
\bibitem{KeCoGo01}
J-M Kelif, Marceau Coupechoux and Philippe Godlewski, 
Spatial Outage Formula for CDMA Networks, VTC Fall 2007
\bibitem{JPe06}
J. Perez, 
Gravity, Dimension, Equilibrium, \& Thermodynamics,
arXiv:astro-ph/0603811v1 Mar 2006
\bibitem{ToTa01}
S.~Toumpis and L.~Tassiulas, 
Packetostatics Deployment of Massively Dense Sensor Networks as an Electrostatics Problem, proc. of INFOCOM, Mar. 2005.
\bibitem{Jac04}
P.~Jacquet,
Geometry of information propagation in massively dense ad hoc networks, Proc. of ACM MobiHoc, May 2004.


\end{thebibliography}
%

\end{document}